\documentclass[aps,pra,twocolumn,superscriptaddress,nofootinbib]{revtex4-2}

\usepackage[braket, qm]{qcircuit}
\usepackage{graphicx}
\usepackage{tikz}
\usetikzlibrary{shapes.geometric, arrows, positioning, calc, shapes.multipart, backgrounds}

\usepackage{subcaption}
\usepackage[T1]{fontenc}
\usepackage{amsmath,amsthm,amssymb,mathrsfs}
\usepackage{indentfirst}
\usepackage{color,xcolor}
\usepackage{verbatim}

\newtheorem{result}{Result}

\newtheorem{remark}{Remark}

\newtheorem{corollary}{Corollary}
\newtheorem{method}{Method}

\newtheorem{definition}{Definition}

\usepackage[colorlinks = true,
            linkcolor = blue,
            urlcolor  = blue,
            citecolor = blue,
            anchorcolor = blue]{hyperref}

\begin{document}

\title{An Efficient Quantum Circuit Construction Method for Mutually Unbiased Bases in $n$-Qubit Systems}

\author{Yu Wang}
\email{wangyu@bimsa.cn}
\affiliation{Beijing Institute of Mathematical Sciences and Applications (BIMSA), Huairou District, Beijing 101408, P. R. China}

\author{Dongsheng Wu}
\email{wudongsheng14@mails.ucas.ac.cn}
\address{Beijing Institute of Mathematical Sciences and Applications (BIMSA), Huairou District, Beijing 101408, P. R. China}
\address{Yau Mathematical Sciences Center, Tsinghua University, Beijing, 100084, China}

\begin{abstract}
Mutually unbiased bases (MUBs) play a crucial role in numerous applications within quantum information science, such as quantum state tomography, error correction, entanglement detection, and quantum cryptography. Utilizing \(2^n + 1\) MUB circuits provides a minimal and optimal measurement strategy for reconstructing all \(n\)-qubit unknown states. It significantly reduces the number of measurements compared to the traditional \(4^n\) Pauli observables, also enhancing the robustness of quantum key distribution (QKD) protocols.
Previous circuit designs that rely on a single generator can result in exponential gate costs for some MUB circuits. In this work, we present an efficient algorithm to generate each of the \(2^n + 1\) quantum MUB circuits on \(n\)-qubit systems within \(O(n^3)\) time. The algorithm features a three-stage structure, and we have calculated the average number of different gates for random sampling. Additionally, we have identified two linear properties: the entanglement part can be directly defined into \(2n - 3\) fixed sub-parts, and the knowledge of \(n\) special MUB circuits is sufficient to construct all \(2^n + 1\) MUB circuits.
This new efficient and simple circuit construction paves the way for the implementation of a complete set of MUBs in diverse quantum information processing tasks on high-dimensional quantum systems.

\end{abstract}

\maketitle

\section{Introduction}

Quantum measurement is the exclusive method for obtaining information about quantum systems, forming a crucial link for understanding microscopic quantum states through empirical observations \cite{braginsky1995quantum}. Projective measurements onto mutually unbiased bases (MUBs) \cite{schwinger1960unitary} are essential and widely utilized in quantum information science.  Preparing an eigenstate of one basis, its distribution is uniform across any other MUB, highlighting their maximal incompatibility and complementarity \cite{bohr1928quantum,maccone2015complementarity,designolle2019quantifying}. MUBs are useful in quantum tomography \cite{ivonovic1981geometrical,wootters1989optimal,adamson2010improving,lima2011experimental}, uncertainty relations \cite{maassen1988generalized,ballester2007entropic,massar2008uncertainty,wu2009entropic}, quantum cryptography \cite{cerf2002security,mafu2013higher,yu2008quantum,casaccino2008extrema,farkas2023mutually}, quantum error correction \cite{calderbank1997quantum,calderbank1998quantum,gottesman1998fault}, and entanglement identification \cite{spengler2012entanglement,giovannini2013characterization,erker2017quantifying,kaniewski2019maximal,tavakoli2021mutually}, to name a few.

Two MUBs can always be constructed in any finite \(d\)-dimensional Hilbert space \cite{schwinger1960unitary} and the upper bound is $d+1$. However, constructing $d+1$ MUBs for each $d$ is remaining an open problem in quantum information theory \cite{horodecki2022five}.  When $d$ is a prime power, $d+1$ MUBs can be constructed \cite{wootters1989optimal}. For dimension $d=6$, strong numerical evidence indicates that there are no four MUBs \cite{butterley2007numerical,bengtsson2007mutually,brierley2009constructing,raynal2011mutually}. Some research focuses on the structure behind complete \( (d+1) \) MUBs sets and incomplete sets \cite{mandayam2013unextendible,goyeneche2013mutually,goyeneche2015mutually}.

To measure the state \( \rho \) using a projective measurement onto one MUB \( \{U_j|k\rangle : k=0,\cdots,d-1\} \), we can apply the unitary \( U_j^{\dag} \) to \( \rho \) and subsequently measure in the computational basis.
We aim to efficiently decompose each of \( 2^n+1 \) circuit for $\{U_j\}$ in \( n \)-qubit systems, starting with circuits involving two MUBs. Even two MUBs usually work in a lot of quantum information tasks \cite{lundeen2011direct,cerf2002security,mafu2013higher}, the \( 2^n+1 \) MUB circuits, together with the computational measurement, are essential as minimal and optimal resources for reconstructing all unknown \( n \)-qubit states \cite{ivonovic1981geometrical,wootters1989optimal,adamson2010improving}. They would be an indispensable component of the verification and certification tools for the future of multi-qubit universal quantum computations. As the number of circuits decreases quadratically compared to \(4^n\) for all Pauli observables.
Additionally, while many \( d \)-dimensional Quantum Key Distribution (QKD) protocols, such as the BB84 protocol, use only two MUBs \cite{bennett2014quantum}, employing \( d+1 \) MUBs enhances QKD robustness, particularly against correlated errors \cite{cerf2002security,wang2020high,ikuta2022scalable}.

Let the first MUB circuit be \( I^{\otimes n} \). The second MUB circuit could be \( H^{\otimes n} \) or a Fourier transformation circuit requiring \( O(n^2) \) gates \cite{nielsen2010quantum}. These circuits are integral to numerous prominent quantum algorithms, including the Deutsch-Jozsa algorithm \cite{deutsch1992rapid,collins1998deutsch,qiu2020revisiting}, Shor's factorization algorithm \cite{shor1994algorithms}, Grover's search algorithm \cite{grover1996fast}, and the HHL algorithm \cite{harrow2009quantum}, among others. However, these two MUB circuits alone cannot generate the complete set of \( 2^n+1 \) MUB circuits directly. Previous works constructed a new second MUB circuit $V$ \cite{chau2005unconditionally,gow2007generation,kern2010complete}. It is interesting to note that by repeating $V$ zero times ($I$), once, twice, and so on up to $2^n$ times, the complete set of $2^n+1$ MUB circuits can be obtained. However, the gate count for some of these circuits could be exponential \cite{seyfarth2011construction,seyfarth2015practical}.

In this work, we introduce a numerical conjecture for identifying both complete and incomplete MUBs using complex Hadamard matrices and diagonal matrices. We choose the $2^n$ MUBs formula obtained by the Galois-Fourier approach \cite{durt2010mutually} to generate the complete  MUBs circuits. Each nontrivial MUB circuit is constructed with the $H^{\otimes n}$ and a diagonal operation, structured as $-H-S-CZ-$. We propose an efficient computational method to decompose each MUB circuit using \( O(n^2) \) gates within \( O(n^3) \) time.  An interesting entanglement structure is presented. We find a linear relation where the knowledge of \( n \) special MUB circuits describes all \( 2^n+1 \) MUB circuits. We calculate the average occurrence of various gate types and analyze the distribution of MUB state coefficients. Finally, we suggest several avenues for further exploration.
 The circuit construction method holds the potential to enhance the utilization of MUBs in the realms of quantum information and quantum computing tasks in the future. And the method could offer deeper insights into MUBs' structural properties.

\section{Preliminaries and a Numerical Method Conjecture}

\begin{definition}[MUB]
A set of two normalized eigenbases \(\{|\psi_j\rangle\}_{j=0}^{d-1}\) and \(\{|\phi_k\rangle\}_{k=0}^{d-1}\) are called mutually unbiased (MU) if the following condition holds for each \(j,k\):
\begin{equation}
    |\langle \psi_j|\phi_k\rangle|^2 = \frac{1}{d}
\end{equation}
\end{definition}

Given a set of \(M\) eigenbases labeled as \(\{\{|\psi_j^k\rangle\}_{j=0}^{d-1} : k=0, \cdots, M-1\}\), if any two bases within this set are MU, then the set is said to contain \(M\) MUBs. For prime power dimensions \(d\), a set containing maximum \(d+1\) MUBs can always be found.

\begin{definition}[Complex Hadamard matrix]
Given a unitary matrix $U$, it is called a complex Hadamard matrix (CHM) if each matrix element $U_{jk}$ satisfies the following condition:
\begin{equation}
    |U_{jk}|^2 = \frac{1}{d}
\end{equation}
\end{definition}

Any CHM can produce the second basis mutually unbiased with computation one $\{|k\rangle\}_{k=0}^{d-1}$. CHMs are used in quantum computing, wireless communications, signal processing, error correction, combinatorial designs, optics, and cryptography to leverage their orthogonal and constant modulus properties \cite{horadam2012hadamard}. The CHM we defined here is \(\frac{1}{\sqrt{d}}\) times of the commonly understood CHM.
With this definition, when \(U_1\) and \(U_2\) are CHMs, both \(U_1 \otimes U_2\) and \(U_1^{\dag}\) are also CHMs.

Fix a basis $\{|\psi_{j}^0\rangle\}_{j=0}^{d-1}$ from a set of MUBs, it corresponds to a unitary operation $I=\sum_{k=0}^{d-1}|\psi_{j}^0\rangle\langle\psi_{j}^0|$.
For any other MUB $\{|\psi_{j}^k\rangle\}_{j=0}^{d-1}$, it corresponds to unitary operation $U_k=\sum_{k=0}^{d-1}|\psi_{j}^k\rangle\langle\psi_{j}^0|$.

\begin{corollary}
Finding a set of $M$ MUBs is equivalent to finding $M-1$ unitary operations $\{U_0=I,U_1,\cdots,U_{M-1}\}$ such that $U_j^{\dag}U_{k}$ is a CHM for all $0\le j<k\le M-1$.
\end{corollary}

\begin{figure}
    \centering
    \includegraphics[width=0.5\linewidth]{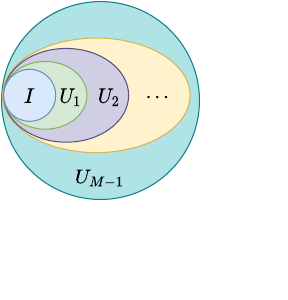}
    \caption{Corollary 1. First, we choose a CHM \( U_1 \). Then we find another CHM \( U_2 \) such that \( U_1^{\dag}U_2 \) is still a CHM. We continue this process until we find the final matrix \( U_{M-1} \) such that \( U_{M-1}^{\dag}U_j \) is a CHM.}
\label{1}
\end{figure}

The corollary yields a numerical method to construct a set of MUBs \cite{bengtsson2007three,mandayam2013unextendible,goyeneche2013mutually,goyeneche2015mutually}, illustrated as Fig.(\ref{1}).
We may as well let the first row of the unitary operations be the real number for the freedom choice of the global phase.
If we have infinite computational resources, the process in Fig.(\ref{1}) will produce all the set of MUBs.

The above method is similar to the brute-force numerical method to construct symmetric informationally complete measurement (SIC-POVM) \cite{renes2004symmetric}, which is to find $d^2$ unit complex vectors $\{|\phi_j\rangle\}_{j=0}^{d^2-1}$ such that the following condition is satisfied for all $0\le j<k\le d^2-1$:
\begin{equation}\label{sic}
    |\langle\phi_j|\phi_k\rangle|^2=1/(d+1)
\end{equation} The existence problems of $d+1$ MUBs and SIC-POVM with $d^2$ elements are identified as the first two open problems in quantum information theory \cite{horodecki2022five}. Zauner's conjecture \cite{zauner1999grundzuge,scott2017sics} simplify the computation process in Eq.(\ref{sic}) by finding one fiducial state $|\phi_0\rangle$. Define $X=\sum_{k=0}^{d-1}|k+1\rangle\langle k|$, $Z=\sum_{k=0}^{d-1}e^{2\pi \sqrt{-1} k/d}|k\rangle\langle k|$. If we can find unit $|\phi_0\rangle$ such that
\begin{equation}\label{Zaunner}
    |\langle \phi_0| X^jZ^k |\phi_0\rangle|^2=1/(d+1)
\end{equation}
 for $j,k=0,\cdots,d-1$ excluding $j=k=0$, the SIC-POVM can be constructed by the Weyl-Heisenberg group acting on the fiducial state. Recently, a relaxed condition has been provided for \(|\phi_0\rangle\) to produce an infinite number of IC-POVMs using the Weyl-Heisenberg group. If it is not satisfied, then there is no fiducial state for SIC-POVM \cite{cao2024dynamical}.

Like the simplification in Eq.(\ref{Zaunner}), we conjecture that the effort to find the maximum number of MUBs, as stated in Corollary 1, can be simplified as follows.
\begin{method}
    In order to construct the unitary operations $\{U_0=I,U_1,\cdots,U_{M-1}\}$ in Corollary 1, we can find a CHM $U_1$ and search out $M-2$ diagonal matrices $D_k$, where $k=1,\cdots,d-2$.
    The diagonal element is chosen from
\begin{equation}\label{D0}
    \{e^{\pi \sqrt{-1}/d}, e^{2\pi \sqrt{-1}/d},\cdots,e^{(2d-1)\pi \sqrt{-1}/d},1\}
\end{equation}
    If the following condition is held,
    \begin{equation}\label{method1}
        U_1^{\dag} D_j^{\dag}D_k U_1 \mbox{~~for ~} 1\le j<k \le M-2
    \end{equation}
    The operations $\{U_0=I,U_1,D_1U_1,\cdots,D_{M-2}U_1\}$ can generate $M$ MUBs.
\end{method}

This method can avoid the knowledge of the mathematical theory in algebra and number theory to construct MUBs.
Once the $d$-level CHM $U_1$ is chosen, the maximum size of the corresponding set of MUBs is fixed. By Eq.(\ref{D0}), we  determine a set $\mathbb{D}_0$ which contains $(2d)^d$ diagonal matrices. Firstly, we can search among $\mathbb{D}_0$ to construct a subset of $\mathbb{D}_1$ where $D_1$ belongs. For each one fixed $D_1$, we can search among $\mathbb{D}_1$ to construct a subset $\mathbb{D}_2$ where $D_2$ belongs. We continue the process until the subset is null.

For $n$-qubit case, $d=2^n$, we let the diagonal elements be $\{\pm1,\pm \sqrt{-1}\}$ and $U_1=H^{\otimes n}$. The first set $\mathbb{D}_0$ contains $4^d$ elements. The numerical experiment indicates that we can always construct a complete set of MUBs using Method 1 for small $n$. Besides, the solutions are not unique.
For example, when $n=1$, the three unitary operations could be $\{I,H,SH\}$ or $\{I,H,S^3H\}$.

However, it seems like things are heading towards two extremes.
The results \cite{seyfarth2011construction,seyfarth2015practical} calculated one generator but some MUB circuits are decomposed of exponential gates. Method 1 fixes one `generator' and costs exponential computations with polynomial decomposed gates for small $n$.

To construct each MUB circuit within polynomial time and using a polynomial number of gates, we turn to the formulas for \(2^n\) nontrivial MUBs. They could be the formula by Wootters and Fields \cite{wootters1989optimal}, the Galois Rings formula \cite{klappenecker2004constructions}, the Galois-Fourier formula \cite{durt2010mutually}, or the method involving the division of \(4^n-1\) Pauli observables \cite{bandyopadhyay2002new,lawrence2002mutually}, and so on. The Galois-Fourier formula \cite{durt2010mutually} directly meets our requirements.

\begin{definition}[Formula by Galois-Fourier method]
Denote $2^n+1$ MUBs as $\{\mathcal{B}_0, \mathcal{C}_0, \mathcal{C}_1, \dots, \mathcal{C}_{2^n-1}\}$, where $\mathcal{B}_0$ is the computational basis $\{|0\rangle, \ldots, |2^n-1\rangle\}$. For $j=0,1,\dots,2^n-1$, basis $\mathcal{C}_j$ contains the following elements \cite[Eq.(2.70)]{durt2010mutually} :
\begin{equation}\label{ekj}
    |e_k^j\rangle=\frac{1}{\sqrt{2^n}}\sum_{l=0}^{2^n -1}|l\rangle(-1)^{k\odot l}\cdot \alpha_l^j
\end{equation}
where $\alpha_l^j=\prod_{s,t=0}^{n-1}\overline{(\sqrt{-1})^{j\odot (l_s\cdot 2^s)\odot(l_t\cdot 2^t)}}$ and $k=0,\cdots,2^n-1$.
\end{definition}

The corresponding MUB operation is $\sum_{k=0}^{2^n-1}|e_k^j\rangle\langle k|=\mathbb{D}\times \mathbb{H}$, where \(\mathbb{H}=\frac{1}{\sqrt{2^n}} \sum_{k,l=0}^{2^n-1} (-1)^{k \odot l^T} |l\rangle \langle k|\) and $\mathbb{D}=\sum_{l=0}^{2^n-1}\alpha_l^j |l\rangle\langle l|$.
Essentially, the definition of the operator \(\mathbb{H}\) is chosen to be a CHM. It can be viewed as the $U_1$ in Method 1, and the circuit structure method lies in the efficient decomposition of \(\mathbb{H}\) and \(\mathbb{D}\).

\subsection{Multiplication $\odot$ in Galois field $GF(2^n)$}

In order to calculate $\odot$ in Eq.(\ref{ekj}),  we first make a brief introduction of $j\odot k$  in the Galois field \( GF(2^n) \) using a polynomial approach.
 We are familiar with the multiplication of integers, $ 3\times 5=15$. When $3,5\in\{0,1,\cdots,7\}$, if we want to guarantee the result $ 3\times 5$ is still in $\in\{0,1,\cdots,7\}$, we can let the result be $7=15 \bmod 8$. Here, the multiplication $j\odot l$ in the Galois field $GF(2^n)$ is similar.

For elements \( j \) and \( k \) in \( [0, 2^n-1] \), they can be expressed in binary form:
\begin{equation}
\left\{
\begin{aligned}
    j &= j_0 + j_1 2^1 + \ldots + j_{n-1} 2^{n-1} \\
    k &= k_0 + k_1 2^1 + \ldots + k_{n-1} 2^{n-1}
\end{aligned}
\right.
\end{equation}
The multiplication \( j \odot k \) involves multiplying these binary forms as polynomials:
\[
\tilde{g}(x) = (j_0 + j_1 x + \ldots + j_{n-1} x^{n-1}) \times (k_0 + k_1 x + \ldots + k_{n-1} x^{n-1})
\]
The product \( \tilde{g}(x) \) is a polynomial of degree up to \( 2n-2 \). To remain within the field \( GF(2^n) \), we reduce this polynomial modulo an irreducible polynomial \( p(x) \) of degree \( n \) over \( \mathbb{F}_2 \):
\[g(x) = \tilde{g}(x) \mod p(x)
\]
The coefficients of \( g(x) \), \( \{g_0, g_1, \ldots, g_{n-1}\} \), define the result of the multiplication in \( GF(2^n) \), represented as:
\begin{equation}\label{odot1}
    j \odot k = g_0 + g_1 2^1 + \ldots + g_{n-1} 2^{n-1}
\end{equation}

This operation is grounded in the structure of the Galois field \( GF(2^n) \), which is constructed by extending \( \mathbb{F}_2 \) with an irreducible polynomial of degree \( n \):
\begin{equation}\label{px}
    p(x) = 1 + a_1x + \ldots + a_{n-1}x^{n-1} + x^n
\end{equation}
The existence of such a polynomial is well-known and there are some fast algorithms generating it \cite{shoup1994fast, couveignes2013fast}.

We denote that the following three expressions are the same. Firstly, the elements of \( GF(2^n) \) are represented as:
\[
l_0 + l_1 x + \ldots + l_{n-1} x^{n-1} \mod p(x)
\]
Secondly, these elements can be viewed as vectors $\{(l_0,\cdots,l_{n-1})\}$ in \( \mathbb{F}_2^n \). Or they are integers $\{l\}$ in the range \( [0, 2^n - 1] \), allowing for operations like addition \( \oplus \) and multiplication \( \odot \).
The multiplication $\odot$ in Eq.(\ref{odot1}) can be simplified:
 \begin{equation}\label{odot}
    j \odot l  = (j \mathcal{M}_0 l^T, \ldots, j \mathcal{M}_{n-1} l^T)
\end{equation}
Here $\mathcal M_0,\mathcal M_1,\dots,\mathcal M_{n-1}$ are invertible matrices in $M_n(\mathbb F_2)$, whose entries are given by the vector forms of $1,x,x^2,\dots,x^{2n-2}$. To be precise, if we write $x^m=\sum_{r=0}^{n-1}(x^m)_r\cdot x^r\pmod{p(x)}$, then
\begin{equation}\label{equ M_r-defn}
    \mathcal M_r=\big((x^{s+t})_r\big)_{1\leqslant s,t\leqslant n-1}.
\end{equation}
The notation $j\mathcal M_rl^T$ means
\begin{equation*}
    j\mathcal M_rl^T=\begin{pmatrix}
        j_0 &j_1 &\cdots &j_{n-1}
    \end{pmatrix}\cdot M_r \cdot
    \begin{pmatrix}
        l_0 &l_1 &\cdots &l_{n-1}
    \end{pmatrix}^T,
\end{equation*}
whose output is an element in $\mathbb F_2=\{0,1\}$.  Detailed discussion about $\odot$ is presented in Appendix A.

With the above notations, we can calculate the  exponential functions
\begin{equation}\label{equ -1-exp}
    (-1)^l:=(-1)^{\sum_{t=0}^{n-1}l_t\cdot 2^t}=(-1)^{l_0}
\end{equation}
and
\begin{equation}\label{equ i-exp}
    (\sqrt{-1})^l:=(\sqrt{-1})^{\sum_{t=0}^{n-1}l_t\cdot 2^t}=(\sqrt{-1})^{l_0+l_1\cdot 2}.
\end{equation}
It is worth noting that $(-1)^{l\oplus j}=(-1)^{l}\cdot(-1)^{j}$, but
\begin{equation}
    (\sqrt{-1})^{l\oplus j}=(\sqrt{-1})^{l}\cdot(\sqrt{-1})^{j} \mbox{~or~} -(\sqrt{-1})^{l}\cdot(\sqrt{-1})^{j}.
\end{equation}

\section{Results and efficient constructions}
We can rearrange the states in Eq.(\ref{ekj}) as follows:
\begin{equation}\label{fjk}
    |f_k^j\rangle=\frac{1}{\sqrt{2^n}}\sum_{l=0}^{2^n -1}|l\rangle(-1)^{k\cdot l^T}\cdot\alpha_l^j
\end{equation}
This is because that states $\{|e_k^j\rangle\}_{k=0}^{2n-1}$ and $\{|f_k^j\rangle\}_{k=0}^{2n-1}$ correspond to the same basis $\mathcal{C}_j$. By Eq.(\ref{odot}), we have $(-1)^{k\odot l}=(-1)^{k\mathcal{M}_0 \cdot l}$. Since $\mathcal{M}_0$ is invertible, when $k$ ranges from $0,\cdots,2^n-1$, the result $k\mathcal{M}_0$ also runs over $0,\cdots,2^n-1$.

Now our task is to implement each circuit $U_j$ for the elements rearranged in basis  $\mathcal{C}_j$:
\begin{equation}
U(j)=\sum_{k=0}^{2^n-1}|f_k^j\rangle\langle k|.
\end{equation}
It is easy to verify that
\begin{equation}
U(j)=\left(\sum_{l=0}^{2^n-1}\alpha_l^j |l\rangle\langle l|\right)\times \left(\frac{1}{\sqrt{2^n}}\sum_{k,l=0}^{2^n-1}(-1)^{k\cdot l^T}|l\rangle\langle k|\right).
\end{equation}

\subsection{Decomposition of $H^{\otimes n}$ part}
The term $\frac{1}{\sqrt{2^n}}\sum_{k,l=0}^{2^n-1}(-1)^{k\cdot l^T}|l\rangle\langle k|$ is exactly the tensor of $n$ Hadamard matrices $H=\frac{1}{\sqrt{2}}\begin{pmatrix}
     1 &1\\
     1& -1
\end{pmatrix}$. The reason for rearrangement is to simplify the circuit decomposition for the CHM part.
\begin{equation}\label{H}
   \frac{1}{\sqrt{2^n}}\sum_{k,l=0}^{2^n-1}(-1)^{k\cdot l^T}|l\rangle\langle k|= H^{\otimes n}
\end{equation}
It is used in the first quantum algorithm, the Deutsch-Jozsa algorithm as referenced in equation (2.55) \cite{nielsen2010quantum}.
It is to produce the maximal balanced state in different quantum algorithms.

The left part $\sum_{l=0}^{2^n-1}\alpha_l^j |l\rangle\langle l|$ is a diagonal operation.
Thus, Method 1 can theoretically yield the complete set of MUBs for \( d = 2^n \). Interestingly, mathematical theories provide a framework that ensures a comprehensive solution for exponential searching.

\subsection{Decomposition of diagonal part}
Now the remaining concern is whether diagonal operation $\sum_{l=0}^{2^n-1}\alpha_l^j |l\rangle\langle l|$ can be decomposed into polynomial gates.
We compute $\alpha_l^j$ to decompose the circuit.
\begin{equation}
    \alpha_l^j=\prod_{s,t=0}^{n-1}\overline{(\sqrt{-1})^{j\odot (l_s\cdot 2^s)\odot(l_t\cdot 2^t)}}
\end{equation}
 We divide the $n^2$ products into two parts: $s=j$ and $s\ne j$.

 When $s=t$, the result of the multiplication part is denoted as
 \begin{equation}\label{equ arj}
     \prod_{r=0}^{n-1}\overline{(\sqrt{-1})^{j\odot (l_r\cdot 2^r)\odot(l_r\cdot 2^r)}}=\prod_{r=0}^{n-1}(\sqrt{-1}\big)^{a_r(j)l_r},
 \end{equation}
where $a_r(j)=0,1,2,3$.

When $s \ne t$, the multiplication $\odot$ can change sequence, $j\odot(l_s2^s)\odot(l_t2^t)=j\odot(l_t2^t)\odot(l_s2^s) $. Thus the result is
\begin{equation}\label{equ bst}
    \prod_{0\leqslant s<t\leqslant n-1} (-1)^{j\odot (l_s\cdot 2^s)\odot(l_t\cdot 2^t)}=\prod_{0\leqslant s<t\leqslant n-1} (-1)^{b_{s,t}(j)l_sl_t},
\end{equation}
where $b_{s,t}(j)=0,1$.

Then we denote $\sum_{l=0}^{2^n-1}\alpha_l^j |l\rangle\langle l|=U_{CZ}(j)\cdot U_{S}(j)$, where
\begin{align}
  U_{S}(j)&= \sum_{l=0}^{2^n-1} \prod_{r=0}^{n-1}(\sqrt{-1}\big)^{a_r(j)l_r} |l\rangle\langle l|, \\
  U_{CZ}(j)&= \sum_{l=0}^{2^n-1} \prod_{0\leqslant s<t\leqslant n-1} (-1)^{b_{s,t}(j)l_sl_t}|l\rangle\langle l|
\end{align}

Thus the structure for the diagonal part is almost given.
Denote the phase gate $S$ as
\begin{equation}
S=|0\rangle\langle0|+\sqrt{-1}|1\rangle\langle1|
\end{equation}
Denote \(CZ(s, t)\) as the \(n\)-qubit Controlled-\(Z\) (\(CZ\)) gate, which acts on qubit \(q_s\) and qubit \(q_t\). Different from the $n$-qubit C-NOT gate, the control and target qubits of the \(CZ\) gate can interchange roles.

By the elementary gates $S$ and $CZ$, we can decompose the diagonal $\sum_{l=0}^{2^n-1}\alpha_l^j |l\rangle\langle l|$ with structure of $-S-CZ-$. Here
\begin{equation} \label{equ Usj}
    U_{S}(j)=S^{a_0(j)}\otimes \cdots\otimes S^{a_{n-1}(j)}
\end{equation}
This means that at the circuit for $\mathcal{C}_j$, we should apply $S$ gate $a_t(j)$ times on qubit $q_t$, where $t=0,\cdots,n-1$. And
\begin{equation} \label{equ UCZj}
    U_{CZ}(j)=\prod_{0\le s<t\le n-1} CZ(s,t)^{b_{s,t}(j)}
\end{equation}
This means we should apply $CZ(s,t)$ gate $b_{s,t}(j)$ times on qubit $q_s$ and $q_t$, where $0\le s<t\le n-1$. Namely, if $b_{s,t}(j)=1$, we add a $CZ$ gate between qubit $q_s$ and $q_t$.

\subsection{Results based on the decomposition}

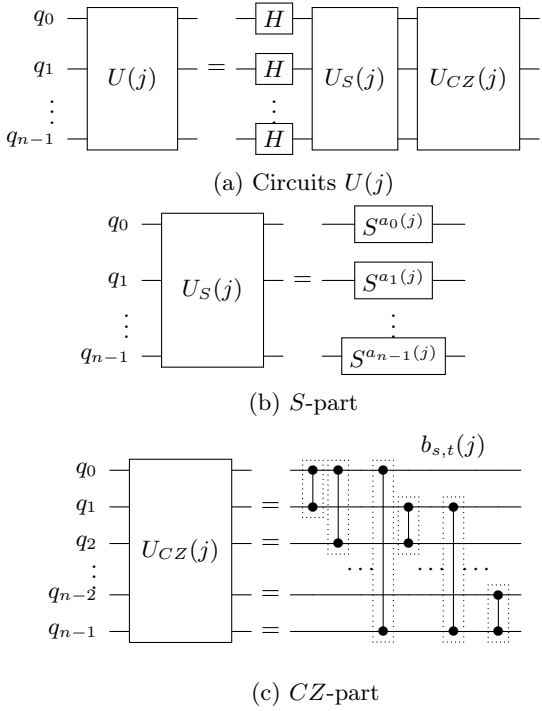
\begin{figure}
  \centering
 \begin{subfigure}{0.4\textwidth}
    \centering
    \begin{minipage}{0.5\textwidth}
        \Qcircuit @C=0.8em @R=0.8em {
            \lstick{q_0}    & \multigate{3}{~U(j)~}& \qw & &  &  \gate{H} & \multigate{3}{U_S(j)}& \multigate{3}{U_{CZ}(j)}& \qw \\
            \lstick{q_1}  &\ghost{~U(j)~} & \qw &=& &      \gate{H} &  \ghost{U_S(j)}&  \ghost{U_{CZ}(j)}& \qw \\
            \lstick{\vdots} &&& &  &\vdots  \\
            \lstick{q_{n-1}}  &\ghost{~U(j)~} & \qw & & & \gate{H} &  \ghost{U_S(j)}&  \ghost{U_{CZ}(j)}& \qw
        }
    \end{minipage}
    \caption{Circuits $U(j)$}
    \label{fig:subfig1}
  \end{subfigure}
~~
  \begin{subfigure}{0.4\textwidth}
    \centering
  \begin{minipage}{1.\textwidth}
  \Qcircuit @C=0.8em @R=0.8em {
\lstick{q_0}    &\multigate{3}{~U_S(j)~}&\qw&   &
&\gate{S^{a_0(j)}} & \qw \\
\lstick{q_1}    &\ghost{~U_S(j)~}&\qw&    =&
&\gate{S^{a_1(j)}} & \qw \\
\lstick{\vdots}  &&&&&\vdots&\\
\lstick{q_{n-1}}    & \ghost{~U_S(j)~}&\qw&   &  & \gate{S^{a_{n-1}(j)}} & \qw }
 \end{minipage}
    \caption{$S$-part}
    \label{fig:subfig2}
  \end{subfigure}
 \\
  \begin{subfigure}{0.5\textwidth}
    \centering

\[
    \Qcircuit @C=0.8em @R=0.6em {
    \lstick{q_0}   &\multigate{5}{U_{CZ}(j)} &\qw &&    &\ctrl{1}   & \ctrl{2} &  \qw & \ctrl{5} &  \qw  &  \qw &  \ustick{b_{s,t}(j)}\qw&  \qw&  \qw&  \qw\\
    \lstick{q_1}    &\ghost{U_{CZ}(j)} &\qw &=&     & \ctrl{0}  & \qw   & \qw  & \qw &\ctrl{1} &  \qw & \ctrl{4}&  \qw&  \qw&  \qw  \\
    \lstick{q_2}    &\ghost{U_{CZ}(j)} &\qw  &=&   & \qw   & \ctrl{0}    & \qw & \qw & \ctrl{0} &  \qw&  \qw&  \qw&  \qw&  \qw\\
    \lstick{\vdots}   &&&& &&&\cdots& &&\cdots &  &\cdots\\
    \lstick{q_{n-2}}  &\ghost{U_{CZ}(j)} &\qw  &=&   &\qw   & \qw    & \qw & \qw& \qw &  \qw&  \qw&  \qw&  \ctrl{1}&  \qw\\
    \lstick{q_{n-1}}  &\ghost{U_{CZ}(j)} &\qw  &=&  &   \qw  & \qw & \qw & \ctrl{0} & \qw & \qw &  \ctrl{0}&  \qw&  \ctrl{0}&  \qw\gategroup{1}{6}{2}{6}{.6em}{.} \gategroup{1}{7}{3}{7}{.6em}{.} \gategroup{1}{9}{6}{9}{.6em}{.} \gategroup{2}{10}{3}{10}{.6em}{.} \gategroup{2}{12}{6}{12}{.6em}{.} \gategroup{5}{14}{6}{14}{.6em}{.}}
 \]
    \caption{$CZ$-part}
    \label{fig:subfig3}
  \end{subfigure}
  \caption{The circuit $U(j), j=0,\dots,2^n-1$. Each  consists of three components: the \(H\)-part, the \(S\)-part \(U_S(j)\), and the \(CZ\)-part \(U_{CZ}(j)\). The coefficients for each part can be directly calculated. The analysis for each component is provided above in Eq.(\ref{H}), Eq.(\ref{equ Usj}), and Eq.(\ref{equ UCZj}).
 }
  \label{fig uj}
\end{figure}

\begin{result}[Circuit structure]
There exists a set of $2^n+1$ MUB circuits constructed with three-stage decomposition.
The circuit for the computational basis $\mathcal{B}_0$ is $I^{\otimes n}$. For $j=0,\dots,2^n-1$, the nontrivial circuit $U(j)$ can be decomposed into the following circuit sequence: $-H-S-CZ-$.
\begin{equation}\label{equ Uj-decom}
    U(j)=U_{CZ}(j)\cdot U_{S}(j)\cdot H^{\otimes n},
\end{equation}

We use Fig.(\ref{fig uj}) to illustrate the Result 1.
\end{result}

As we can see in Fig.(\ref{fig uj}), the \(H\)- and \(S\)-parts are straightforward to implement since they do not involve entanglement. The entangled \(CZ\)-part, however, contains at most \(\frac{n(n-1)}{2}\) different types of \(CZ(s,t)\) operations, where \(0 \le s < t \le n-1\). Next, we will discuss the \(CZ\)-part with an analysis of the coefficient \(b_{s,t}(j)\).

\begin{result}[Entanglement structure]
For each nontrivial MUB circuit $U(j)$, where $j\in\{0,1,\dots, 2^n-1\}$, there are two facts about the entanglement part.
\begin{itemize}
    \item [(I)] If gate $CZ(s,t)$ appears at $U(j)$, then $CZ(s',t')$ should also appear.
    Here $0\leqslant s<t\leqslant n-1, 0\leqslant s'<t'\leqslant n-1$ and $s+t=s'+t'$.
    \item [(II)] We can define $2n-3$ fixed $CZ$ sub-parts without calculation of $\odot$:
    \begin{equation}\label{czm}
       CZ(m)=  \prod_{\stackrel{0\leqslant s<t\leqslant n-1}{s+t=m}}CZ(s,t),
    \end{equation}
    where $m=0,1,\cdots,2n-2$.
    Each $-CZ-$ part $U_{CZ}(j)$ in Fig.(\ref{fig:subfig3}) is a linear combination of Eq.(\ref{czm}).
    Precisely, denote $b_{s+t}(j):=b_{s,t}(j)\in \{0,1\}$. Then
    \begin{equation}
        U_{CZ}(j)=\prod_{m=1}^{2n-3}\bigg(CZ(m)\bigg)^{b_{m}(j)}.
    \end{equation}
    In other words, by preparing the $2n-3$ $CZ$ sub-parts $CZ(m)$, we can combine them to generate each $U_{CZ}(j)$ in Fig.(\ref{fig:subfig2}).
    \end{itemize}
\end{result}

\emph{Proof.} Recall that $b_{s,t}(j)\in\{0,1\}$ and
\begin{equation*}
    (-1)^{b_{s,t}(j)l_sl_t}=(-1)^{j\odot (l_s\cdot 2^s)\odot(l_t\cdot 2^t)}
\end{equation*}
for any $l_s,l_t\in\{0,1\}$. The exponential on the right-hand side is equal to
\begin{equation*}
    j\odot(l_s\cdot x^s)\odot(l_t\cdot x^t)=j\odot(x^{s+t})\cdot (l_sl_t).
\end{equation*}
By Eq.\eqref{equ -1-exp},
\begin{equation*}
\begin{split}
    (-1)^{b_{s,t}(j)\cdot(l_sl_t)}=&(-1)^{j\odot(x^{s+t})\cdot (l_sl_t)}\\
    =&(-1)^{j\mathcal M_0\cdot (x^{s+t})^T\cdot(l_sl_t)}.
\end{split}
\end{equation*}
So
\begin{equation}\label{equ bst-formula}
    b_{s,t}(j)=j\mathcal M_0\cdot (x^{s+t})^T.
\end{equation}
Hence
\begin{equation*}
    b_{s,t}(j)=j\mathcal M_0\cdot (x^{s+t})^T=j\mathcal M_0\cdot (x^{s'+t'})^T=b_{s',t'}(j),
\end{equation*}
which proves part (I); and part (II) arises from Eq.\eqref{equ UCZj} and part (I).  \qed

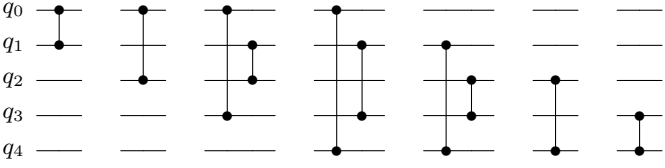
\begin{figure}[!htb]
  \centering
    \begin{minipage}[t]{0.4\textwidth}
    \vspace{0.5cm} 
    \Qcircuit @C=0.8em @R=1.2em {
    \lstick{q_0}  &\ctrl{1} & \qw   && &\ctrl{2} & \qw  && &    \ctrl{3} & \qw      &\qw  &&  & \ctrl{4} & \qw & \qw && &\qw &\qw &\qw  && &\qw &\qw && &\qw &\qw\\
    \lstick{q_1}  &\ctrl{0} & \qw   && &\qw      & \qw  && &    \qw      & \ctrl{1} &\qw  &&  & \qw &\ctrl{2} & \qw && &\ctrl{3} &\qw &\qw && &\qw &\qw && &\qw &\qw\\
    \lstick{q_2}  &\qw      & \qw   && &\ctrl{0} & \qw  && &    \qw      & \ctrl{0} &\qw  &&  & \qw & \qw & \qw && &\qw &\ctrl{1} &\qw && &\ctrl{2} &\qw && &\qw &\qw\\
    \lstick{q_3}  &\qw      & \qw   && &\qw      & \qw  && &    \ctrl{0} & \qw      &\qw  &&  & \qw & \ctrl{0} & \qw && &\qw &\ctrl{0} &\qw && &\qw &\qw && &\ctrl{1} &\qw\\
    \lstick{q_4}  &\qw    & \qw   && &\qw      & \qw  && &  \qw     & \qw      &\qw  &&  &\ctrl{0} & \qw & \qw
    && &\ctrl{0} &\qw &\qw && &\ctrl{0} &\qw && &\ctrl{0} &\qw
    }
    \end{minipage}
    \vspace{0.5cm}
  \caption{Seven $CZ$ sub-parts by Eq.(\ref{czm}) for 5-qubit.
  Each $U_{CZ}(j)$ in Fig.(\ref{fig:subfig3}) is a linear combination of them.
  These parts $CZ(m)$ are pre-fixed without the need for multiplication $\odot$.
  For example, we consider the third circuit for  $CZ(3)$. As $3=0+3=1+2$, $CZ(3)=CZ(0,3)\cdot CZ(1,2)$.}
  \label{CZ(m)}
\end{figure}

We make a brief illustration for 5-qubit $CZ(m)$ in Fig.(\ref{CZ(m)}).

Now that we have a clear understanding of each of the three components for all circuits \(\{U(j)\}\), we can deduce the time complexity required to obtain them.

\begin{result}[Time complexities]
Given an input qubit number $n$ and a random $j\in\{0,1,\dots,2^n-1\}$, the time complexity to generate the circuit $U(j)$ is $O(n^3)$.
\end{result}

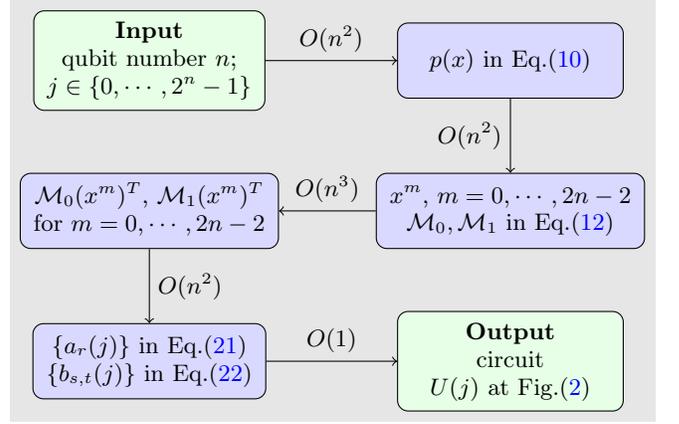
\begin{figure}
\centering
\begin{tikzpicture}[node distance=1.8cm, startstop/.style={rectangle, rounded corners, minimum width=3cm, minimum height=1cm,text centered, draw=black, fill=white}, background rectangle/.style={fill=gray!20}, show background rectangle]

\def\startcolor{green!10}
\def\stopcolor{green!10}
\def\intermediatecolor{blue!15}

\node (start) [startstop, fill=\startcolor] {\begin{tabular}{c}
        \textbf{Input} \\
        qubit number $n$; \\
        $j\in\{0,\cdots,2^n-1\}$
    \end{tabular}};
\node (proc1) [startstop, right of=start, xshift=3cm, fill=\intermediatecolor] {  $p(x)$ in Eq.(\ref{px})};
\node (proc2) [startstop, below of=proc1, yshift=-0.2cm, fill=\intermediatecolor] {\begin{tabular}{c}
        $x^m$, $m=0,\cdots,2n-2$\\
        $\mathcal{M}_0, \mathcal{M}_1$ in Eq.(\ref{equ M_r-defn})
    \end{tabular}};
\node (proc3) [startstop, left of=proc2, xshift=-3cm, fill=\intermediatecolor] { \begin{tabular}{c}
        $\mathcal{M}_0 (x^m)^{T}$, $\mathcal{M}_1 (x^m)^{T}$\\
        for $m=0,\cdots,2n-2$
    \end{tabular}};
\node (proc4) [startstop, below of=proc3, yshift=-0.2cm, fill=\intermediatecolor] {\begin{tabular}{c}
        $\{a_r(j)\}$ in Eq.(\ref{equ arj})\\
        $\{b_{s,t}(j)\}$ in Eq.(\ref{equ bst})
    \end{tabular}};
\node (stop) [startstop, right of=proc4, xshift=3cm, fill=\stopcolor] {\begin{tabular}{c}
        \textbf{Output} \\
           circuit  \\
        $U(j)$ at Fig.(\ref{fig uj})
    \end{tabular}};

\draw [->] (start) -- node[above] {$O(n^2)$} (proc1);
\draw [->] (proc1) -- node[left] {$O(n^2)$}(proc2);
\draw [->] (proc2) -- node[above] {$O(n^3)$} (proc3);
\draw [->] (proc3) -- node[right] {$O(n^2)$}(proc4);
\draw [->] (proc4) -- node[above] {$O(1)$} (stop);

\end{tikzpicture}
\caption{The procedure of generating the circuit $U(j)$. For $n$-qubit systems, the time complexity is $O(n^3)$ for every $j=0,1,\dots,2^n-1$. After obtaining the coefficients $\{a_r(j)\},\{b_{s,t}(j)\}$, the circuit $U_j$ is directly determined, which is expressed succinctly as $O(1)$.
}
\label{fig process}
\end{figure}

\emph{Analysis.} The procedure to generate $U(j)$ (see Fig. \ref{fig process}) and the time complexities of each step are as summarized as follows.
\begin{itemize}
    \item [Step 1.] It requires $O(n^2)$ operations to generate an irreducible polynomial $p(x)$ in Eq.(\ref{px}) according to \cite{shoup1994fast, couveignes2013fast}. The presence of numerous zeros within the set ${a_1,\cdots,a_{n-1}}$ can simplify subsequent calculations. For example, $p(x)=x^9+x+1$ when $n=9$. Such polynomials for $2\le n \le 10000$ are listed in \cite{seroussi1998table}.
    \item [Step 2.] For an arbitrary $x^m, m=0,1,\dots,2m-2$, it takes at most $O(n)$ operations to obtain its vector form according to Eq.\eqref{app-equ x1-x2-xn} and Eq.\eqref{app-equ xm-rec} in the Appendix. Hence it takes $O(n^2)$ operations to compute the vector forms of all $x^m, m=0,1,\dots,2n-2$. Meanwhile, we obtain $\mathcal M_0$ and $\mathcal M_1$ automatically by Eq.\eqref{app-equ M_r-defn}.
    \item [Step 3.] In order to calculate all parts about $\odot$, it takes $O(n^2)$ operations to obtain each of the vectors $\mathcal M_0(x^m)^T$ and $\mathcal M_1(x^m)^T, m=0,1,\dots, 2n-2$. So it requires $O(n^3)$ operations in total to get all these vectors.
    \item [Step 4.] With the vectors $\mathcal M_0(x^m)^T$ and $\mathcal M_1(x^m)^T, m=0,1,\dots, 2n-2$ in hand, it requires $O(n)$ operations to compute each of the coefficients $a_r(j)$ and $b_{s,t}(j)=b_{s+t}(j)$. Note that by Result 2, not $O(n^2)$ coefficients about $b_{s,t}(j)$, we only need to compute
    \begin{equation}
\left\{
\begin{aligned}
   & a_0(j),a_1(j),\dots,a_{n-1}(j); \\
   & b_1(j),b_2(j),\dots,b_{2n-3}(j).
\end{aligned}
\right.
\end{equation}
 So it takes at most $O(n^2)$ operations to obtain all these coefficients, which are sufficient to generate $U(j)$.
\end{itemize}
To sum up, given qubit number $n$ and index $j$, it takes at most $O(n^3)$ operations to generate the circuit $U(j)$ in Fig.(\ref{fig uj}). \qed

\begin{remark}
Given an input qubit number $n$. If one wants to generate a sequence of circuits $U(j^{(1)})$, $U(j^{(2)})$, $\dots$, $U(j^{(T)})$ simultaneously, the time complexity is $O(n^3+Tn^2)$, not $O(Tn^3)$. This is because we only need to compute the vectors $\mathcal M_0(x^m)^T$ and $\mathcal M_1(x^m)^T, m=0,1,\dots, 2n-2$ once.
\end{remark}

A linear relationship has been observed for the entanglement part \(CZ\), where each \(U_{CZ}(j)\) is a linear combination of \(CZ(m)\). Furthermore, a surprising result reveals that there are \(n\) special MUB circuits that form the `core' of the \(2^n\) nontrivial MUB circuits \(\{U_j : j = 0, \dots, 2^n-1\}\). We refer to this as the second linear relation.

\begin{result}[Linear Relation]
There is a `linear' relation between the \(2^n\) MUB circuits \(U(j)\). Specifically, given the knowledge of \(n\) circuits \(U(2^0), U(2^1),\cdots, U(2^{n-1})\), any circuit \(U(j)\) can be deduced for \(0 \le j \le 2^n-1\).
\end{result}

\emph{Analysis.}
For each \(j\), it has a binary expression given by \(j = j_0 2^0 + \cdots + j_{n-1} 2^{n-1}\). To determine the circuit for \(U(j)\), we only need to determine the \(S\) part \(U_{S}(j)\) and the \(CZ\) part \(U_{CZ}(j)\), as the \(H\) part is fixed.

Assume the coefficients $b_{s,t}(2^0)$, $b_{s,t}(2^1)$, $\dots$, $b_{s,t}(2^{n-1})$ and $a_r(2^0)$, $a_r(2^1)$,$\dots$, $a_r(2^{n-1})$ are already known. We will show that for any $0\leqslant j\leqslant n-1$, the coefficients $b_{s,t}(j)$ and $a_r(j)$ in Fig.(\ref{fig uj}) can be obtained by the two linear formulas Eq.\eqref{equ bst-decom} and Eq.\eqref{equ arj-decom}.

Fix a pair $(s,t)$ and write $j=(j_0,j_1,\dots,j_{n-1})$. Then Eq.\eqref{equ bst} implies
\begin{equation}\label{equ bst-decom}
    b_{s,t}(j)=\sum_{u=0}^{n-1}j_u\cdot b_{s,t}(2^u)\pmod{2}.
\end{equation}

For the $-CZ-$ part of $U(j)$,
\begin{equation}
  U_{CZ}(j)=U_{CZ}(2^0)^{j_0}\times \cdots  \times U_{CZ}(2^{n-1})^{j_{n-1}}
\end{equation}
For example, if $j=3=(1,1,0,0)$, we just need to combine $U_{CZ}(2^0)$ and $U_{CZ}(2^1)$  to obtain $U_{CZ}(j)$. If $CZ(s,t)$ appears twice, we can eliminate to $I$ as $CZ^2(s,t)=I$.

Now we consider the $S$-part.
It is a bit complicated as $a_r(j)\in \{0,1,2,3\}$.
Recall that the value $a_r(j)$ is determined by
\begin{equation*}
   (\sqrt{-1}\big)^{a_r(j)l_r}=\overline{(\sqrt{-1})^{j\odot (l_r\cdot 2^r)\odot(l_r\cdot 2^r)}}
\end{equation*}
By Eq.\eqref{equ i-exp}, the right-hand side is equal to
\begin{equation*}
    \overline{(\sqrt{-1})^{j\odot x^{2r}\cdot l_r}}=\overline{(\sqrt{-1})^{(j\mathcal M_0\cdot (x^{2r})^T+j\mathcal M_1\cdot (x^{2r})^T\cdot 2)\cdot l_r}}.
\end{equation*}
Here we should be careful about the conjugate operation $\overline{x}$ for complex number $x$.
If we denote $(a,b)=(j\mathcal M_0\cdot(x^{2r})^T,j\mathcal M_1\cdot(x^{2r})^T)\in\mathbb F_2^2$, then
\begin{equation}\label{equ arj-formula}
    a_r(j)=\begin{cases}
        0,&~\text{if}~(a,b)=(0,0)\\
        1,&~\text{if}~(a,b)=(1,1)\\
        2,&~\text{if}~(a,b)=(0,1)\\
        3,&~\text{if}~(a,b)=(1,0)
    \end{cases}
\end{equation}

It should be noted that
\begin{equation*}
    a_r(j\oplus j')\ne a_r(j)+a_r(j') \pmod 4
\end{equation*}
in general.
This means that we can not sum up the $S$ gate numbers and take modulo 4 like $CZ$ part.
However, if we define
\begin{equation*}
    \tau(a):=\begin{cases}
        (0,0),\quad&\text{if}~a=0\\
        (1,1),\quad&\text{if}~a=1\\
        (0,1),\quad&\text{if}~a=2\\
        (1,0),\quad&\text{if}~a=3
    \end{cases}
\end{equation*}
then the vector
\begin{equation*}
    \tau(a_r(j))=(j\mathcal M_0\cdot(x^{2r})^T, j\mathcal M_1\cdot(x^{2r})^T)
\end{equation*}
is linear with respect to $j$. So for $j=(j_0,\cdots,j_{n-1})$,
\begin{equation}\label{equ arj-decom}
    \tau(a_r(j))=\sum_{u=0}^{n-1}j_u\cdot\tau(a_r(2^u))\pmod 2.
\end{equation}
Since $\tau$ defines a bijection between $\{0,1,2,3\}$ and $\{0,1\}^2$, the coefficients $a_r(2^0), a_r(2^1),\cdots, a_r(2^{n-1})$ can determine any $a_r(j)$ by Eq.\eqref{equ arj-decom}.

 We use an example to show how to obtain the  $-S-$ part of $U(j)$.
 If $j=3=(1,1,0,0)$, we should check the $-S-$ part of circuit $U(2^0)$ and $U(2^1)$. For qubit $q_2$, assume that $S$ and $S^3$ appears at $U(2^0)$ and $U(2^1)$ respectively. We know $\tau(a_2(3))=(1,1)+(1,0) \pmod 2=(0,1)$ by Eq.(\ref{equ arj-decom}). Thus $a_2(3)=2$. Then gate $S^2$ appears at qubit $q_2$ of $U(3)$.
\qed

After obtaining the entanglement structure and understanding the linear behaviors of the \(2^n\) nontrivial MUB circuits from a global perspective, a key question arises: how many gates (upper bound or average number) will appear if we randomly select a MUB circuit? This understanding is crucial for estimating the experimental cost associated with using randomized MUB circuits in quantum information tasks.

\begin{remark}[Maximal gate number]
For each circuit \(U(j)\), where \(j = 0, 1, \dots, 2^n-1\), the count of \(S\) gates is \(\sum_{r=0}^{n-1} a_r(j)\), and the count of \(CZ\) gates is \(\sum_{0 \le s < t \le n-1} b_{s,t}(j)\).

For a fixed \(j\), the maximal number of \(S\) gates is \(3n\) if \(a_r(j) = 3\) for all \(r = 0, \dots, n-1\), and the maximal number of \(CZ\) gates is \(\frac{n(n-1)}{2}\) if \(b_{s,t}(j) = 1\) for all \(0 \le s < t \le n-1\).

Adding the \(n\) \(H\) gates, the maximal number of gates in any circuit \(U(j)\) is \(\frac{n^2 + 7n}{2}\).
\end{remark}

The value \(\frac{n^2 + 7n}{2}\) provides the upper bound for the gate cost in one randomly selected MUB circuit. Next, we estimate the average gate cost when randomly selecting a large number of MUB circuits.

\begin{result}[Average gates counting]
 Aggregating across the sets $\{U(j)\}_{j=0}^{2^n-1}$, the average number of $S$ gates is $3n/2$, the average number of $CZ$ gates is $(n^2-n)/4$, and the number of $CZ$ gates of distance $u$ (represented as $CZ(s,t)$ with $t-s=u$) is $(n-u)/2$.
\end{result}

\emph{Analysis.} Let us look at the $CZ$ gates first. Fixed a pair $(s,t), 0\leqslant s<t\leqslant n-1$. Since $\mathcal M_0$ is invertible, the vector $\mathcal M_0\cdot (x^{s+t})^T$ is nonzero. As $j$ varies from $0$ to $2^n-1$, half of the values of $b_{s,t}(j)=j\mathcal M_0\cdot (x^{s+t})^T$ are $1$, while the other half are $0$.
So the total number of $CZ$ gates and the total number of $CZ$ gates with distance $u$ are
\begin{equation*}
\begin{split}
       \sum_{j=0}^{2^n-1}\sum_{0\leqslant s<t\leqslant n-1}b_{s,t}(j)=&\sum_{0\leqslant s<t\leqslant n-1}\sum_{j=0}^{2^n-1}b_{s,t}(j)\\
       =&\frac{n^2-n}{2}\cdot 2^{n-1}=2^n\cdot\frac{n^2-n}{4}
\end{split}
\end{equation*}
and
\begin{equation*}
\begin{split}
     &\sum_{j=0}^{2^n-1}\sum_{\stackrel{0\leqslant s<t\leqslant n-1}{t-s=u}}b_{s,t}(j)=\sum_{\stackrel{0\leqslant s<t\leqslant n-1}{t-s=u}}\sum_{j=0}^{2^n-1}b_{s,t}(j)\\
     =&\sum_{\stackrel{0\leqslant s<t\leqslant n-1}{t-s=u}} 2^{n-1}=(n-u)\cdot 2^{n-1}=2^n\cdot \frac{n-u}{2}.
\end{split}
\end{equation*}

Next we consider the average number of $S$ gates in $U(j)$. For any fixed $r\in\{0,1,\dots, n-1\}$, the value $j\odot (x^{2r})$ runs over $0,1,\dots,2^n-1$ when $j$ varies from $0$ to $2^n-1$. So for each value in $(a,b)\in\{0,1\}\times\{0,1\}$, a quarter of the pairs $(j\mathcal M_0\cdot(x^{2r})^T, j\mathcal M_1\cdot(x^{2r})^T)$ take the value $(a,b)$. Hence
\begin{equation*}
\begin{split}
    \sum_{j=0}^{2^n-1}\sum_{r=0}^{n-1}a_r(j)=&\sum_{r=0}^{n-1}\sum_{j=0}^{2^n-1}a_r(j)\\
    =&n\times(0+1+2+3)\times 2^{n-2}=2^n\cdot\frac{3}{2}n.
\end{split}\qed
\end{equation*}

After calculating the average number of gates, we can apply a similar approach to analyze the distributions of coefficients across all \(2^n \times 2^n\) nontrivial MUB states. This analysis is useful for estimating the experimental costs associated with using randomized MUB states in quantum information tasks.

We may as well define the $l$-th component of $|f_k^j\rangle$ as
\begin{equation}
    \langle l|f_k^j \rangle=\sqrt{2^{-n}} c(j,k,l)
\end{equation}
Interestingly, $c(j,k,l)\in\{\pm 1,\pm\sqrt{-1}\}$. These coefficients $c(j,k,l)$ exhibit some regularities.

\begin{result}[Coefficient distribution]
If $l=0$, $c(j,k,0)$ is always $1$. Consider the nontrivial case when $l>0$.
\begin{itemize}
    \item [(i)] Fix the basis $\mathcal{C}_j$.
    The values in the sequence $c(j,0,l)$, $c(j,1,l)$, $\dots$, $c(j,2^n-1,l)$ are evenly divided, with half being $1$ and the other half being $-1$, or half being $\sqrt{-1}$ and the other half being $-\sqrt{-1}$;
    \item [(ii)] The $4^n$ values $c(j,k,l), j,k=0,1,\dots,2^n-1$ are evenly distributed, with a quarter being $1$, another quarter being $-1$, a quarter being $\sqrt{-1}$, and the remaining quarter being $-\sqrt{-1}$.
\end{itemize}
\end{result}

\emph{Analysis.} (i) Recall that
\begin{equation*}
    c(j,k,l)=(-1)^{k\cdot l^T}\cdot \prod_{0\leqslant s,t\leqslant n-1}\overline{(\sqrt{-1})^{j\odot (l_s\cdot 2^s)\odot(l_t\cdot 2^t)}}.
\end{equation*}
Since $l>0$, half of the values $(-1)^{k\cdot l^T}, k=0,1,\dots,2^n-1$, are $1$ and the other half are $-1$. The remaining product on the rightmost relies solely on $j$ and $l$, taking values from $\{\pm1,\pm\sqrt{-1}\}$. If this product is $\pm 1$, then half of the values $c(j,0,l),\dots,c(j,2^n-1,l)$ are $1$ and the other half are $-1$; if this product is $\pm \sqrt{-1}$, then half of the values are $\sqrt{-1}$ and the other half are $-\sqrt{-1}$.

(ii) Let $j,j'\in GF(2^n)$. By the definition of exponential with base $\sqrt{-1}$ (see Eq.\eqref{equ -1-exp}), we have
\begin{equation*}
    \sqrt{-1}^{j}\cdot\sqrt{-1}^{j'}=\pm\sqrt{-1}^{j\oplus j'}
\end{equation*}
Hence
\begin{equation*}
    \prod_{0\leqslant s,t\leqslant n-1}\big(\sqrt{-1}\big)^{j\odot(l_s\cdot 2^s)\odot(l_t\cdot 2^t)}=\pm \big(\sqrt{-1}\big)^{j\odot l}.
\end{equation*}
Since $j\odot l$ runs over the values $0,1,\dots,2^n-1$ as $j$ ranges from $0$ to $2^n-1$, we see that half of the values $\big(\sqrt{-1}\big)^{j\odot l}, j=0,1,\dots,2^n-1$, are in $\{\pm 1\}$, and the other half are in $\{\pm\sqrt{-1}\}$. Combining with part (i), we see that (ii) holds.

\section{Conclusion and discussions}

Efficient decomposition of \(2^n + 1\) MUB circuits is crucial for enhancing performance across various domains of quantum information and computation, including the minimal and optimal reconstruction of all unknown \(n\)-qubit states \cite{adamson2010improving}, improving the robustness of QKD protocols \cite{ikuta2022scalable}, reducing the sample complexity of randomized measurements to extract specific information of unknown states \cite{wang2024classical}, and various verification protocols, among others. In this work, we achieve the decomposition of each MUB circuit in \(O(n^3)\) time and with \(O(n^2)\) elementary gates using the Galois-Fourier formula. A detailed discussion of the circuits is given. The circuit structure \( -H-S-CZ- \) represents the simplest part of all Clifford circuits generated by \(\{H, S, CZ \text{ (or CNOT)}\}\). We identify a pre-fixed \(2n-3\) entanglement substructure and a linear relation that the core of \(2^n\) nontrivial MUB circuits is actually the $n$ special ones. We also analyze the average gate costs and distribution of coefficients for all MUB states during randomized measurements.

There are several interesting problems with the MUB structure for future work.

\begin{figure}[htp]
    \centering
    \includegraphics[width=0.35\linewidth]{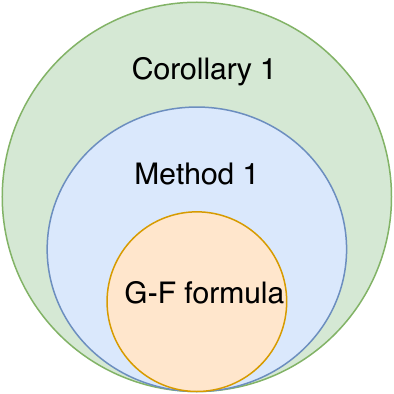}
    \caption{Comparision of construction method.}
    \label{relation}
\end{figure}

Firstly, when \(d\) is a prime power, Fig.(\ref{relation}) lists different methods to produce a maximum of \(d+1\) MUBs.
Corollary 1 can output all MUBs theoretically.
In cases where \(d\) is not a prime power, if Corollary 1 produces at most \(M\) MUBs, can Method 1 always generate \(M\) MUBs? Alternatively, given \(M\) MUBs produced by Corollary 1, can Method 1 generate \(M\) MUBs using certain mathematical techniques?

Secondly, how does the initial CHM \( U_1 \) affect the number of MUBs generated by Method 1? When \( U_1 \) is chosen as the global Fourier transform \( F \), theoretically, only two MUBs, \(\{I, F\}\), exist for $d=6$ \cite{grassl2004sic, brierley2009constructing}. While changing to others, three MUBs can be found. Characterizing all possible initial \( U_1 \) choices in Method 1 or the Galois-Fourier formula to achieve the maximum number of MUBs is an intriguing task.
Developing a picture language for the circuits to efficiently verify Eq.(\ref{method1}) is also an interesting pursuit.

Thirdly, will there be other MUB circuit structures? Shortly after we utilized random circuits for states \( |f_k^j\rangle \) in Eq.(\(\ref{fjk}\)) for classical shadow tomography in the first edition, Zhang et al. \cite{zhang2024minimal} independently derived the three-stage structure of \(2^n + 1\) MUB circuits, as well as the structure of the \(CZ\) components. It seems that the circuits are constructed for the original states \( |e_k^j\rangle \). An important perspective is that MUB circuits are a subset of Clifford circuits, enabling methods designed for Clifford circuits \cite{gottesman1997stabilizer, aaronson2004improved, maslov2018shorter, bravyi2021hadamard} to encompass all MUB formulas. These techniques could be applied not only to the MUB circuits for the Galois-Fourier formula \cite{zhang2024minimal} but also to other constructions \cite{wootters1989optimal, klappenecker2004constructions, bandyopadhyay2002new, lawrence2002mutually}.
Besides, the output circuits generated by the Galois-Fourier formula can vary depending on the choice of different irreducible polynomials \( p(x) \) in Eq.(\ref{px}). Exploring the underlying structure behind these circuits is also intriguing.

Lastly, simplifying our circuits further is an intriguing task. With infinite ways to decompose each \( U(j) \), optimizing gate count, circuit depth, or introducing ancillas during the compilation of these \( U(j) \) presents an interesting challenge. Reexamining the entanglement structure and linear relations then could provide new insights for MUB circuits.

\textbf{Acknowledgements---}
We thank the helpful discussions with Jinsong Wu, Ruijie Xu and Zhuo Chen.
This work received support from the National Natural Science Foundation of China through Grants No. 62001260 and No. 42330707, as well as from the Beijing Natural Science Foundation under Grant No. Z220002.

\textbf{Author contribution---}
Y. W. conceived the idea of this paper. D. W. performed the calculations over the Galois field. Y. W. designed the circuits based on these calculations. Both Y. W. and D. W. wrote the manuscript.

\clearpage

\appendix

\section{Operations in $GF(2^n)$}
We investigate the rules of operations in $GF(2^n)$ with respect to the vector form in detail.

Let $j=(j_0,\cdots,j_{n-1})$ and $l=(l_0,\dots,l_{n-1})$. Then
\begin{equation*}
\begin{split}
   j\oplus l=&\sum_{r=0}^{n-1}j_rx^r+\sum_{r=0}^{n-1}l_rx^r=\sum_{r=0}^{n-1}(j_r+l_r)x^r\\
   =&(j_0+l_0,\cdots,j_{n-1}+l_{n-1}).
\end{split}
\end{equation*}
The multiplication $k\odot l$ is equal to
\begin{equation}\label{equ jcdotl}
  \big(\sum_{r=0}^{n-1}j_rx^r\big)\cdot\big(\sum_{r=0}^{n-1}l_rx^r\big)=\sum_{s,t=0}^{n-1}j_s l_t x^{s+t}.
\end{equation}
Write
\begin{equation*}
x^m\equiv \sum_{r=0}^{n-1}(x^m)_rx^r\pmod{p(x)}.
\end{equation*}
Hence the $r$-th component in the vector form of Eq.\eqref{equ jcdotl} is equal to
\begin{equation*}
    \sum_{s,t=0}^{n-1} j_s\cdot (x^{s+t})_r\cdot l_t=\begin{pmatrix}
        j_0 &j_1 &\cdots &j_{n-1}
\end{pmatrix}
\cdot \mathcal M_r\cdot
\begin{pmatrix}
        l_0\\
        l_1\\
        \vdots\\
        l_{n-1}
\end{pmatrix},
\end{equation*}
where
\begin{equation}\label{app-equ M_r-defn}
\begin{split}
\mathcal M_r:=&\big((x^{s+t})_r\big)_{1\leqslant s,t\leqslant n-1}\\
=&\begin{pmatrix}
    (x^0)_r & (x^1)_r & \cdots & (x^{n-1})_r\\
    (x^1)_r & (x^2)_r & \cdots & (x^n)_r \\
    \vdots  & \vdots  & \ddots & \vdots \\
    (x^{n-1})_r & (x^n)_r & \cdots & (x^{2n-2})_r
\end{pmatrix}
\end{split}
\end{equation}
for $r=0,\dots,2^n-1$. Then the multiplication rule in $GF(2^n)$ could be written as
\begin{equation*}
    j\odot l=(j\mathcal M_0l^T,j\mathcal M_1l^T,\cdots, j\mathcal M_{n-1}l^T).
\end{equation*}

We illustrate a way to compute the vector representation of $x^m$. Recall that $p(x)=1+a_1x+\cdots+a_{n-1}x^{n-1}+x^n$. For powers not greater than $n$, we have
\begin{equation}\label{app-equ x1-x2-xn}
\begin{split}
    x^0=&(1,0,\cdots,0),~~x^1=(0,1,\cdots,0),\\
    x^{n-1}=&(0,0,\cdots,1),~~x^n=(1,a_1,\cdots,a_{n-1}),
\end{split}
\end{equation}
and we can compute high powers via the recursive relation:
\begin{equation}\label{app-equ xm-rec}
\begin{split}
    x^{m+1}=&x^m\cdot x=\big(\sum_{r=0}^{n-1}(x^m)_r\cdot x^r\big)\cdot x\\
    =&\sum_{r=0}^{n-2}(x^m)_r\cdot x^{r+1}+(x^m)_{n-1}\cdot x^n\\
    =&(0, (x^m)_0,(x^m)_1,\cdots,(x^m)_{n-2})\\
    &+(x^m)_{n-1}\cdot(1,a_1,a_2,\cdots,a_{n-1}).
\end{split}
\end{equation}

We remark here that the matrices $\mathcal M_r\in  M_{\mathbb F_2}(n), r=0,1,\dots,2^n-1$, are invertible. This is because for any $j\ne 0$, there exists $l$ such that $j\odot l=x^r$ (since $GF(2^n)$ is a field); in particular, $j\mathcal M_r l^T=1$. So $j\mathcal M_r\ne 0$ if $j\ne 0$, which implies $\mathcal M_r$ is invertible.

\section{MUBs circuits for $n=1,2,3$ and the verifications of MU}

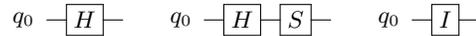
\begin{figure}[!htb]
  \[
  \Qcircuit @C=0.8em @R=0.8em {
\lstick{q_0}    & \gate{H} &\qw}
 ~~~~~~~~~~
  \Qcircuit @C=0.8em @R=0.8em {
\lstick{q_0} & \gate{H} & \gate{S} &\qw  }
 ~~~~~~~~~~
   \Qcircuit @C=0.8em @R=0.8em {
\lstick{q_0} & \gate{I} &\qw  }
 ~~~~~~~~~~
 \]
  \caption{1-qubit MUBs circuits. From left to right, the circuits correspond to the eigenbases of Pauli observable $X,Y,Z$ respectively.}
  \label{fig 1-qubit}
\end{figure}

For $n=1$, the three MUBs are the eigenstates of Pauli observable $X,Y,Z$. The MUBs circuits are as in Fig.\ref{fig 1-qubit}.

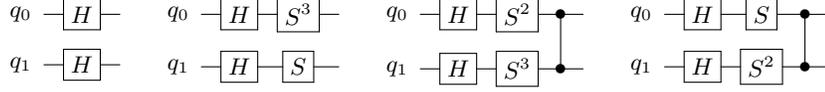
\begin{figure}[!htb]
  \[
  \Qcircuit @C=0.8em @R=0.8em {
\lstick{q_0}    & \gate{H} &\qw \\
\lstick{q_1}    & \gate{H} &\qw}
 ~~~~~~~~~~
  \Qcircuit @C=0.8em @R=0.8em {
\lstick{q_0}    & \gate{H}  & \gate{S^3}&\qw \\
\lstick{q_1}    & \gate{H}  & \gate{S}   &\qw}
 ~~~~~~~~~~
   \Qcircuit @C=0.8em @R=0.8em {
\lstick{q_0}    & \gate{H}  & \gate{S^2} &\ctrl{1}&\qw \\
\lstick{q_1}    & \gate{H}  & \gate{S^3} &\ctrl{0}&\qw}
 ~~~~~~~~~~
    \Qcircuit @C=0.8em @R=0.8em {
\lstick{q_0}    & \gate{H}  & \gate{S}&\ctrl{1}&\qw \\
\lstick{q_1}    & \gate{H}  & \gate{S^2}&\ctrl{0}&\qw}
 ~~~~~~~~~~
 \]
  \caption{2-qubit MUBs circuits.}
  \label{fig 2-qubit}
\end{figure}

For $n=2$, we choose the irreducible polynomial $p(x)$ as $x^2+x+1$. The $2^2$ MUBs circuits without $I^{\otimes n}$ are depicted in Fig.\ref{fig 2-qubit}.

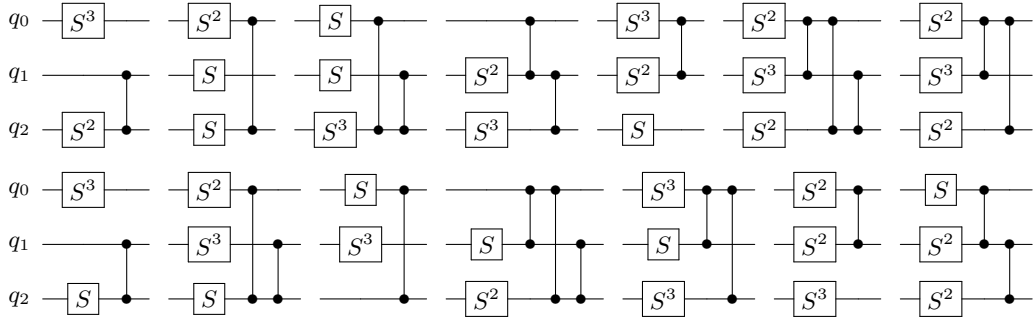
\begin{figure}[!htb]
\[
 \Qcircuit @C=0.8em @R=0.8em {
\lstick{q_0}    & \gate{S^3}&\qw&\qw &~~~~ & \gate{S^2}&\ctrl{2}&\qw&~~~~& \gate{S} &\ctrl{2}&\qw&\qw&~~~~& \qw&\ctrl{1}&\qw &\qw &~~~~& \gate{S^3}&\ctrl{1}&\qw &~~~~& \gate{S^2}&\ctrl{1}&\ctrl{2}&\qw &\qw&~~~~& \gate{S^2}&\ctrl{1}&\ctrl{2}&\qw\\
\lstick{q_1}    & \qw   &\ctrl{1}&\qw &~~~~ & \gate{S}   &\qw&\qw&~~~~& \gate{S}  &\qw&\ctrl{1}&\qw&~~~~& \gate{S^2}   &\ctrl{0}&\ctrl{1}&\qw&~~~~& \gate{S^2}   &\ctrl{0}&\qw&~~~~& \gate{S^3}   &\ctrl{0}&\qw&\ctrl{1}&\qw&~~~~& \gate{S^3}   &\ctrl{0}&\qw&\qw\\
\lstick{q_2}    & \gate{S^2} &\ctrl{0}&\qw&~~~~& \gate{S} &\ctrl{0}&\qw&~~~~& \gate{S^3} &\ctrl{0}&\ctrl{0}&\qw&~~~~&\gate{S^3} &\qw&\ctrl{0}&\qw&~~~~& \gate{S} &\qw&\qw&~~~~& \gate{S^2} &\qw&\ctrl{0}&\ctrl{0}&\qw&~~~~& \gate{S^2} &\qw&\ctrl{0}&\qw}
\]
\[
 \Qcircuit @C=0.8em @R=0.8em {
\lstick{q_0}    & \gate{S^3}&\qw&\qw&~~~~  & \gate{S^2}&\ctrl{2}&\qw&\qw&~~~~  & \gate{S} &\ctrl{2} &\qw &~~~~  & \qw      &\ctrl{1}&\ctrl{2}&\qw&\qw&~~~~  & \gate{S^3}&\ctrl{1}&\ctrl{2}&\qw &~~~~  & \gate{S^2}&\ctrl{1}&\qw&~~~~ & \gate{S}   &\ctrl{1}&\qw     &\qw\\
\lstick{q_1}    & \qw   &\ctrl{1}&\qw&~~~~  & \gate{S^3}   &\qw&\ctrl{1}&\qw& ~~~~  & \gate{S^3}  &\qw &\qw& ~~~~  & \gate{S} &\ctrl{0}&\qw&\ctrl{1}&\qw& ~~~~  & \gate{S}&\ctrl{0}&\qw&\qw& ~~~~  & \gate{S^2}&\ctrl{0}&\qw& ~~~~ & \gate{S^2} &\ctrl{0}&\ctrl{1}&\qw\\
\lstick{q_2}    & \gate{S} &\ctrl{0}&\qw & ~~~~ & \gate{S} &\ctrl{0}&\ctrl{0}&\qw& ~~~~  & \qw &\ctrl{0} &\qw& ~~~~ & \gate{S^2}&\qw&\ctrl{0}&\ctrl{0}&\qw& ~~~~ & \gate{S^3} &\qw     &\ctrl{0}&\qw& ~~~~ & \gate{S^3} &\qw&\qw& ~~~~ & \gate{S^2} &\qw     &\ctrl{0}&\qw}
\]
\caption{3-qubit MUBs circuits (the first line is related with $p_1(x)$, the second line is related with $p_2(x)$. We omit the $H$-part. From the left to right, they correspond to   $U(1),U(2),\dots,U(7)$, respectively.}
\label{fig 3-qubit}
\end{figure}

For $n=3$, we choose $p_1(x)=x^3+x+1$ and $p_2(x)=x^3+x^2+1$. The two sets of $2^3$ circuits are depicted in Fig.\ref{fig 3-qubit}.

\begin{remark}
Notice that the entanglement structures given by $p_1(x)$ and $p_2(x)$ in Fig.\ref{fig 3-qubit} coincide. Namely, their $CZ$ parts are the same up to a rearrangement. This is just a coincidence. In fact, for $n=4$, the three irreducible polynomials $x^4+x+1$, $x^4+x^3+1$ and $x^4+x^3+x^2+x+1$ generate three different entanglement structures.
It is interesting to explore the types of polynomials that might yield reduced levels of entanglement. Additionally, exploring the performance of different types of $2^n+1$ MUBs across various information processing tasks is an intriguing aspect worth exploring in the future.
\end{remark}

\onecolumngrid

\begin{remark}
    To confirm whether circuits are MU, we can use the condition in Eq.(\ref{method1}).
Some picture language would help us decrease the verification cost.
\end{remark}

\begin{figure}[!htb]
  \[
  \Qcircuit @C=0.8em @R=0.8em {
\lstick{q_0}    & \gate{H}    & \gate{S} & \gate{H} &\qw }
 ~~~~~~~~~~
    \Qcircuit @C=0.8em @R=0.8em{
\lstick{q_0}    & \gate{H}    &\qw }
 ~~~~~~~~~~
   \Qcircuit @C=0.8em @R=0.8em {
\lstick{q_0}    & \gate{H}    & \gate{S}   &\qw }
 ~~~~~~~~~~
 \]
 \caption{Checking for 1-qubit. The above three circuits are CHMs.} \label{check1}
\end{figure}
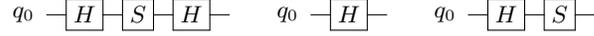

For example, we can easily check that the first circuit in Fig.(\ref{check1}) is a CHM, which confirms that the first two circuits for 1-qubit are MU.

\begin{figure}[!htb]
  \[
  \Qcircuit @C=0.8em @R=0.8em {
\lstick{q_0}    & \gate{H}  & \gate{S^3}  & \gate{H}&\qw \\
\lstick{q_1}    & \gate{H}  & \gate{S}   & \gate{H}  &\qw}
 ~~~~~~~~~~
   \Qcircuit @C=0.8em @R=0.8em {
\lstick{q_0}    & \gate{H}  & \gate{S} &\ctrl{1}& \gate{H} &\qw \\
\lstick{q_1}    & \gate{H}  & \gate{I} &\ctrl{0}& \gate{H} &\qw}
 \]
  \caption{Checking for 2-qubit. We briefly list two checking circuits.}
  \label{check2}
\end{figure}
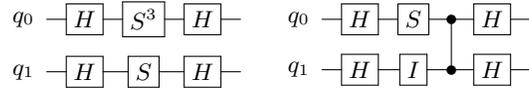

Using the CHM circuit depicted in Fig.(\ref{check1}), we can verify that the first circuits in 2-qubit systems are mutually unbiased (MU). For instance, the circuit \((HS^3H)\otimes(HSH)\) serves as a CHM, which is depicted in the first circuit in Fig.(\ref{check2}). The tensor of two CHMs is still a CHM.
Similarly, to confirm that the second and third circuits are MU, we can verify that \(H^{\otimes 2}(S \otimes I)CZ(1,2) H^{\otimes 2}\) also acts as a CHM.
The checking circuit for the last two MUB circuits is equivalent to the first two MUB circuits in 2-qubit case.

\begin{thebibliography}{10}

\bibitem{braginsky1995quantum}
Vladimir~B Braginsky and Farid~Ya Khalili.
\newblock {\em Quantum measurement}.
\newblock Cambridge University Press, 1995.

\bibitem{schwinger1960unitary}
Julian Schwinger.
\newblock Unitary operator bases.
\newblock {\em Proceedings of the National Academy of Sciences},
  46(4):570--579, 1960.

\bibitem{bohr1928quantum}
Niels Bohr et~al.
\newblock {\em The quantum postulate and the recent development of atomic
  theory}, volume~3.
\newblock Printed in Great Britain by R. \& R. Clarke, Limited, 1928.

\bibitem{maccone2015complementarity}
Lorenzo Maccone, Dagmar Bru{\ss}, and Chiara Macchiavello.
\newblock Complementarity and correlations.
\newblock {\em Physical review letters}, 114(13):130401, 2015.

\bibitem{designolle2019quantifying}
S{\'e}bastien Designolle, Paul Skrzypczyk, Florian Fr{\"o}wis, and Nicolas
  Brunner.
\newblock Quantifying measurement incompatibility of mutually unbiased bases.
\newblock {\em Physical review letters}, 122(5):050402, 2019.

\bibitem{ivonovic1981geometrical}
ID~Ivonovic.
\newblock Geometrical description of quantal state determination.
\newblock {\em Journal of Physics A: Mathematical and General}, 14(12):3241,
  1981.

\bibitem{wootters1989optimal}
William~K Wootters and Brian~D Fields.
\newblock Optimal state-determination by mutually unbiased measurements.
\newblock {\em Annals of Physics}, 191(2):363--381, 1989.

\bibitem{adamson2010improving}
RBA Adamson and Aephraim~M Steinberg.
\newblock Improving quantum state estimation with mutually unbiased bases.
\newblock {\em Physical review letters}, 105(3):030406, 2010.

\bibitem{lima2011experimental}
Gustavo Lima, Leonardo Neves, R~Guzm{\'a}n, Esteban~S G{\'o}mez, WAT Nogueira,
  Aldo Delgado, A~Vargas, and Carlos Saavedra.
\newblock Experimental quantum tomography of photonic qudits via mutually
  unbiased basis.
\newblock {\em Optics Express}, 19(4):3542--3552, 2011.

\bibitem{maassen1988generalized}
Hans Maassen and Jos~BM Uffink.
\newblock Generalized entropic uncertainty relations.
\newblock {\em Physical review letters}, 60(12):1103, 1988.

\bibitem{ballester2007entropic}
Manuel~A Ballester and Stephanie Wehner.
\newblock Entropic uncertainty relations and locking: Tight bounds for mutually
  unbiased bases.
\newblock {\em Physical Review A}, 75(2):022319, 2007.

\bibitem{massar2008uncertainty}
Serge Massar and Philippe Spindel.
\newblock Uncertainty relation for the discrete fourier transform.
\newblock {\em Physical review letters}, 100(19):190401, 2008.

\bibitem{wu2009entropic}
Shengjun Wu, Sixia Yu, Klaus M{\o}lmer, et~al.
\newblock Entropic uncertainty relation for mutually unbiased bases.
\newblock {\em Physical Review A}, 79(2):022104, 2009.

\bibitem{cerf2002security}
Nicolas~J Cerf, Mohamed Bourennane, Anders Karlsson, and Nicolas Gisin.
\newblock Security of quantum key distribution using d-level systems.
\newblock {\em Physical review letters}, 88(12):127902, 2002.

\bibitem{mafu2013higher}
Mhlambululi Mafu, Angela Dudley, Sandeep Goyal, Daniel Giovannini, Melanie
  McLaren, Miles~J Padgett, Thomas Konrad, Francesco Petruccione, Norbert
  L{\"u}tkenhaus, and Andrew Forbes.
\newblock Higher-dimensional orbital-angular-momentum-based quantum key
  distribution with mutually unbiased bases.
\newblock {\em Physical Review A}, 88(3):032305, 2013.

\bibitem{yu2008quantum}
I-Ching Yu, Feng-Li Lin, and Ching-Yu Huang.
\newblock Quantum secret sharing with multilevel mutually (un) biased bases.
\newblock {\em Physical Review A}, 78(1):012344, 2008.

\bibitem{casaccino2008extrema}
Andrea Casaccino, Ernesto~F Galvao, and Simone Severini.
\newblock Extrema of discrete wigner functions and applications.
\newblock {\em Physical Review A}, 78(2):022310, 2008.

\bibitem{farkas2023mutually}
M{\'a}t{\'e} Farkas, J{\k{e}}drzej Kaniewski, and Ashwin Nayak.
\newblock Mutually unbiased measurements, hadamard matrices, and superdense
  coding.
\newblock {\em IEEE Transactions on Information Theory}, 2023.

\bibitem{calderbank1997quantum}
A~Robert Calderbank, Eric~M Rains, Peter~W Shor, and Neil~JA Sloane.
\newblock Quantum error correction and orthogonal geometry.
\newblock {\em Physical Review Letters}, 78(3):405, 1997.

\bibitem{calderbank1998quantum}
A~Robert Calderbank, Eric~M Rains, Peter~M Shor, and Neil~JA Sloane.
\newblock Quantum error correction via codes over gf (4).
\newblock {\em IEEE Transactions on Information Theory}, 44(4):1369--1387,
  1998.

\bibitem{gottesman1998fault}
Daniel Gottesman.
\newblock Fault-tolerant quantum computation with higher-dimensional systems.
\newblock In {\em NASA International Conference on Quantum Computing and
  Quantum Communications}, pages 302--313. Springer, 1998.

\bibitem{spengler2012entanglement}
Christoph Spengler, Marcus Huber, Stephen Brierley, Theodor Adaktylos, and
  Beatrix~C Hiesmayr.
\newblock Entanglement detection via mutually unbiased bases.
\newblock {\em Physical Review A}, 86(2):022311, 2012.

\bibitem{giovannini2013characterization}
D~Giovannini, J~Romero, Jonathan Leach, A~Dudley, A~Forbes, and Miles~J
  Padgett.
\newblock Characterization of high-dimensional entangled systems via mutually
  unbiased measurements.
\newblock {\em Physical review letters}, 110(14):143601, 2013.

\bibitem{erker2017quantifying}
Paul Erker, Mario Krenn, and Marcus Huber.
\newblock Quantifying high dimensional entanglement with two mutually unbiased
  bases.
\newblock {\em Quantum}, 1:22, 2017.

\bibitem{kaniewski2019maximal}
J{k{e}}drzej Kaniewski, Ivan {\v{S}}upi{\'c}, Jordi Tura, Flavio Baccari,
  Alexia Salavrakos, and Remigiusz Augusiak.
\newblock Maximal nonlocality from maximal entanglement and mutually unbiased
  bases, and self-testing of two-qutrit quantum systems.
\newblock {\em Quantum}, 3:198, 2019.

\bibitem{tavakoli2021mutually}
Armin Tavakoli, M{\'a}t{\'e} Farkas, Denis Rosset, Jean-Daniel Bancal, and
  Jedrzej Kaniewski.
\newblock Mutually unbiased bases and symmetric informationally complete
  measurements in bell experiments.
\newblock {\em Science advances}, 7(7):eabc3847, 2021.

\bibitem{horodecki2022five}
Pawe{\l} Horodecki, {\L}ukasz Rudnicki, and Karol {\.Z}yczkowski.
\newblock Five open problems in quantum information theory.
\newblock {\em PRX Quantum}, 3(1):010101, 2022.

\bibitem{butterley2007numerical}
Paul Butterley and William Hall.
\newblock Numerical evidence for the maximum number of mutually unbiased bases
  in dimension six.
\newblock {\em Physics Letters A}, 369(1-2):5--8, 2007.

\bibitem{bengtsson2007mutually}
Ingemar Bengtsson, Wojciech Bruzda, {\AA}sa Ericsson, Jan-{\AA}ke Larsson,
  Wojciech Tadej, and Karol {\.Z}yczkowski.
\newblock Mutually unbiased bases and hadamard matrices of order six.
\newblock {\em Journal of mathematical physics}, 48(5), 2007.

\bibitem{brierley2009constructing}
Stephen Brierley and Stefan Weigert.
\newblock Constructing mutually unbiased bases in dimension six.
\newblock {\em Physical Review A}, 79(5):052316, 2009.

\bibitem{raynal2011mutually}
Philippe Raynal, Xin L{\"u}, and Berthold-Georg Englert.
\newblock Mutually unbiased bases in six dimensions: The four most distant
  bases.
\newblock {\em Physical Review A}, 83(6):062303, 2011.

\bibitem{mandayam2013unextendible}
Prabha Mandayam, Somshubhro Bandyopadhyay, Markus Grassl, and William~K
  Wootters.
\newblock Unextendible mutually unbiased bases from pauli classes.
\newblock {\em arXiv preprint arXiv:1302.3709}, 2013.

\bibitem{goyeneche2013mutually}
Dardo Goyeneche.
\newblock Mutually unbiased triplets from non-affine families of complex
  hadamard matrices in dimension 6.
\newblock {\em Journal of Physics A: Mathematical and Theoretical},
  46(10):105301, 2013.

\bibitem{goyeneche2015mutually}
Dardo Goyeneche and Santiago Gomez.
\newblock Mutually unbiased bases with free parameters.
\newblock {\em Physical Review A}, 92(6):062325, 2015.

\bibitem{lundeen2011direct}
Jeff~S Lundeen, Brandon Sutherland, Aabid Patel, Corey Stewart, and Charles
  Bamber.
\newblock Direct measurement of the quantum wavefunction.
\newblock {\em Nature}, 474(7350):188--191, 2011.

\bibitem{bennett2014quantum}
Charles~H Bennett and Gilles Brassard.
\newblock Quantum cryptography: Public key distribution and coin tossing.
\newblock {\em Theoretical computer science}, 560:7--11, 2014.

\bibitem{wang2020high}
Fumin Wang, Pei Zeng, Jiapeng Zhao, Boris Braverman, Yiyu Zhou, Mohammad
  Mirhosseini, Xiaoli Wang, Hong Gao, Fuli Li, Robert~W Boyd, et~al.
\newblock High-dimensional quantum key distribution based on mutually partially
  unbiased bases.
\newblock {\em Physical Review A}, 101(3):032340, 2020.

\bibitem{ikuta2022scalable}
Takuya Ikuta, Seiseki Akibue, Yuya Yonezu, Toshimori Honjo, Hiroki Takesue, and
  Kyo Inoue.
\newblock Scalable implementation of (d+ 1) mutually unbiased bases for
  d-dimensional quantum key distribution.
\newblock {\em Physical Review Research}, 4(4):L042007, 2022.

\bibitem{nielsen2010quantum}
Michael~A Nielsen and Isaac~L Chuang.
\newblock {\em Quantum computation and quantum information}.
\newblock Cambridge university press, 2010.

\bibitem{deutsch1992rapid}
David Deutsch and Richard Jozsa.
\newblock Rapid solution of problems by quantum computation.
\newblock {\em Proceedings of the Royal Society of London. Series A:
  Mathematical and Physical Sciences}, 439(1907):553--558, 1992.

\bibitem{collins1998deutsch}
David Collins, KW~Kim, and WC~Holton.
\newblock Deutsch-jozsa algorithm as a test of quantum computation.
\newblock {\em Physical Review A}, 58(3):R1633, 1998.

\bibitem{qiu2020revisiting}
Daowen Qiu and Shenggen Zheng.
\newblock Revisiting deutsch-jozsa algorithm.
\newblock {\em Information and Computation}, 275:104605, 2020.

\bibitem{shor1994algorithms}
Peter~W Shor.
\newblock Algorithms for quantum computation: discrete logarithms and
  factoring.
\newblock In {\em Proceedings 35th annual symposium on foundations of computer
  science}, pages 124--134. Ieee, 1994.

\bibitem{grover1996fast}
Lov~K Grover.
\newblock A fast quantum mechanical algorithm for database search.
\newblock In {\em Proceedings of the twenty-eighth annual ACM symposium on
  Theory of computing}, pages 212--219, 1996.

\bibitem{harrow2009quantum}
Aram~W Harrow, Avinatan Hassidim, and Seth Lloyd.
\newblock Quantum algorithm for linear systems of equations.
\newblock {\em Physical review letters}, 103(15):150502, 2009.

\bibitem{chau2005unconditionally}
Hoi~Fung Chau.
\newblock Unconditionally secure key distribution in higher dimensions by
  depolarization.
\newblock {\em IEEE Transactions on Information Theory}, 51(4):1451--1468,
  2005.

\bibitem{gow2007generation}
Rod Gow.
\newblock Generation of mutually unbiased bases as powers of a unitary matrix
  in 2-power dimensions.
\newblock {\em arXiv preprint math/0703333}, 2007.

\bibitem{kern2010complete}
Oliver Kern, Kedar~S Ranade, and Ulrich Seyfarth.
\newblock Complete sets of cyclic mutually unbiased bases in even prime-power
  dimensions.
\newblock {\em Journal of Physics A: Mathematical and Theoretical},
  43(27):275305, 2010.

\bibitem{seyfarth2011construction}
Ulrich Seyfarth and Kedar~S Ranade.
\newblock Construction of mutually unbiased bases with cyclic symmetry for
  qubit systems.
\newblock {\em Physical Review A}, 84(4):042327, 2011.

\bibitem{seyfarth2015practical}
U~Seyfarth, LL~Sanchez-Soto, and G~Leuchs.
\newblock Practical implementation of mutually unbiased bases using quantum
  circuits.
\newblock {\em Physical Review A}, 91(3):032102, 2015.

\bibitem{durt2010mutually}
Thomas Durt, Berthold-Georg Englert, Ingemar Bengtsson, and Karol
  {\.Z}yczkowski.
\newblock On mutually unbiased bases.
\newblock {\em International journal of quantum information}, 8(04):535--640,
  2010.

\bibitem{horadam2012hadamard}
Kathy~J Horadam.
\newblock {\em Hadamard matrices and their applications}.
\newblock Princeton university press, 2012.

\bibitem{bengtsson2007three}
Ingemar Bengtsson.
\newblock Three ways to look at mutually unbiased bases.
\newblock In {\em AIP Conference Proceedings}, volume 889, page~40. AIP, 2007.

\bibitem{renes2004symmetric}
Joseph~M Renes, Robin Blume-Kohout, Andrew~J Scott, and Carlton~M Caves.
\newblock Symmetric informationally complete quantum measurements.
\newblock {\em Journal of Mathematical Physics}, 45(6):2171--2180, 2004.

\bibitem{zauner1999grundzuge}
Gerhard Zauner.
\newblock Grundz{\"u}ge einer nichtkommutativen designtheorie.
\newblock {\em Ph. D. dissertation, PhD thesis}, 1999.

\bibitem{scott2017sics}
Andrew~J Scott.
\newblock Sics: Extending the list of solutions.
\newblock {\em arXiv preprint arXiv:1703.03993}, 2017.

\bibitem{cao2024dynamical}
Meng Cao, Tenghui Deng, and Yu~Wang.
\newblock Dynamical quantum state tomography with time-dependent channels.
\newblock {\em Journal of Physics A: Mathematical and Theoretical},
  57(21):215301, 2024.

\bibitem{klappenecker2004constructions}
Andreas Klappenecker and Martin R{\"o}tteler.
\newblock Constructions of mutually unbiased bases.
\newblock In {\em Finite Fields and Applications: 7th International Conference,
  Fq7, Toulouse, France, May 5-9, 2003. Revised Papers}, pages 137--144.
  Springer, 2004.

\bibitem{bandyopadhyay2002new}
Bandyopadhyay, Boykin, Roychowdhury, and Vatan.
\newblock A new proof for the existence of mutually unbiased bases.
\newblock {\em Algorithmica}, 34:512--528, 2002.

\bibitem{lawrence2002mutually}
Jay Lawrence, {\v{C}}aslav Brukner, and Anton Zeilinger.
\newblock Mutually unbiased binary observable sets on n qubits.
\newblock {\em Physical Review A}, 65(3):032320, 2002.

\bibitem{shoup1994fast}
Victor Shoup.
\newblock Fast construction of irreducible polynomials over finite fields.
\newblock {\em Journal of Symbolic Computation}, 17(5):371--391, 1994.

\bibitem{couveignes2013fast}
Jean-Marc Couveignes and Reynald Lercier.
\newblock Fast construction of irreducible polynomials over finite fields.
\newblock {\em Israel Journal of Mathematics}, 194:77--105, 2013.

\bibitem{seroussi1998table}
Gadiel Seroussi.
\newblock {\em Table of low-weight binary irreducible polynomials}.
\newblock Hewlett-Packard Laboratories, 1998.

\bibitem{wang2024classical}
Yu~Wang and Wei Cui.
\newblock Classical shadow tomography with mutually unbiased bases.
\newblock {\em Physical Review A}, 109(6):062406, 2024.

\bibitem{grassl2004sic}
Markus Grassl.
\newblock On sic-povms and mubs in dimension 6.
\newblock {\em arXiv preprint quant-ph/0406175}, 2004.

\bibitem{zhang2024minimal}
Qingyue Zhang, Qing Liu, and You Zhou.
\newblock Minimal-clifford shadow estimation by mutually unbiased bases.
\newblock {\em Physical Review Applied}, 21(6):064001, 2024.

\bibitem{gottesman1997stabilizer}
Daniel Gottesman.
\newblock {\em Stabilizer codes and quantum error correction}.
\newblock California Institute of Technology, 1997.

\bibitem{aaronson2004improved}
Scott Aaronson and Daniel Gottesman.
\newblock Improved simulation of stabilizer circuits.
\newblock {\em Physical Review A}, 70(5):052328, 2004.

\bibitem{maslov2018shorter}
Dmitri Maslov and Martin Roetteler.
\newblock Shorter stabilizer circuits via bruhat decomposition and quantum
  circuit transformations.
\newblock {\em IEEE Transactions on Information Theory}, 64(7):4729--4738,
  2018.

\bibitem{bravyi2021hadamard}
Sergey Bravyi and Dmitri Maslov.
\newblock Hadamard-free circuits expose the structure of the clifford group.
\newblock {\em IEEE Transactions on Information Theory}, 67(7):4546--4563,
  2021.

\end{thebibliography}
\end{document}